\documentclass[aps,prd,preprint,groupedaddress]{revtex4}
\usepackage{epsf,epsfig,graphics}

\begin{document}

\def\bdm{\begin{equation}}
\def\edm{\end{equation}}
\def\bea{\begin{eqnarray}}
\def\eea{\end{eqnarray}}
\def\ba{\begin{array{}}}
\def\ea{\end{array}}
\def\phir{\phi\hspace{-1.2mm}/}
\def\rhor{\rho\hspace{-1.2mm}/}
\def\GFr{G\hspace{-1.8mm}/_F}
\def\taur{\tau\hspace{-1.2mm}/}
\def\Es{E_{\star}}

\preprint{PITT-0111}

\title{Nonperturbative bound on high multiplicity cross sections in
        $\phi^4_3$ from lattice simulation}



\author{Y.-Y.Charng}
\email[]{charng@phys.sinica.edu.tw, charng@phyast.pitt.edu}
\affiliation{Institute of Physics,Academia Sinica\\Taipei, Taiwan 115,
Republic of China}

\author{R.S.Willey}
\email[]{willey@pitt.edu}
\affiliation{Department of Physics, University of Pittsburgh\\
Pittsburgh, PA 15260}


\date{\today}

\begin{abstract}

We have looked for evidence of large cross sections at large multiplicities
in weakly coupled scalar field theory in three dimensions. We use spectral 
function sum rules to derive bounds on total cross sections where the sum 
can be expresed in terms of a quantity which can be measured by Monte Carlo
simulation in Euclidean space. We  find that high multiplicity cross sections
remain small for energies and multiplicities for which large effects had been  suggested.

\end{abstract}


\maketitle


\section{\label{sec:level1}\protect\\Introduction}


   In 1990, Ringwald \cite{Ri:nphb350} and Espinosa \cite{Es:nphb343} presented 
a calculation of possible 
instanton induced baryon number violation accompanied by enhanced production 
of a large number of Higgs and W,Z bosons at high energy($E \sim n\; m/g$,    where $n$
 is the number of outgoing particles of mass m, and g is a coupling constant
of the theory). 
This  has called attention to a more general and profound problem of
qantum field theory. The problem is: for a given renormalized quantum field 
theory with weak renormalized coupling constant, what is the nature of the
multiparticle production in that theory.

  It has long been known that the perturbation-theory expansion in a theory 
with weak coupling can fail in high orders because of the factorial growth of
the coefficients in the series \cite{Zj:phyrp70}. So this problem can 
not be solved 
perturbatively because the perturbation series  diverges
rapidly at high order. 
In fact, independent of the instanton calculations,
Goldberg \cite{Go:phyletb246}, Cornwall\cite{Co:phyletb243} and, later 
Voloshin\cite{Vo:phyletb293}, Argyres, Kleiss, and Papadopoulos
\cite{AKG:nphb341},and Brown\cite{Br:prd46}
pointed out that in $\phi^4$ scalar field theory the contributions of just the
tree graphs to the multiparticle production amplitudes gives the $k!g^k$ 
behavior. 

These complete tree graph calculations lead to a dramatic exponential growth
of the rates for few $\rightarrow$ many processes as the total energy E and number of
final particles n increase together, and E exceeds some critical value of order $E \ge n m/g$. Something must turn this exponential growth off as it saturates
unitarity bounds, but the phenomenological consequences could still be 
spectacular. Or perhaps the tree graph calculation is simply misleading -
the dramatic exponential turn on only occurs for values of $n \sim 1/g$
for which the perturbation series diverges. 
Perhaps the answer is given just by taking the first few (small) terms
of an asymptotic perturbation series with a small coupling constant.
On the other extreme, A number of semiclassical calculations have suggested
that these high multiplicity events are exponentially suppressed. 
\cite{Rub:Rev}

 It was found by Mawhinney and Willey \cite{Wm:prl} that a nonperturbative
approach could be made through  Monte Carlo simulation on a lattice. It was
shown that using the analytic continuation from Euclidean lattice theory
to Minkowski space quantum field theory, one could turn spectral function
sum rules into an upper bound on inclusive multiparticle production.
The program was carried through for $\phi^4$ theory in $1+1$ dimension
because,as explained below, large lattices are required, and we only 
had the computing power for large two-dimensional ($256^2$) lattices. No
signal for significant highmultiplicity processes was seen.
 However, the $1+1$ 
dimensional theory is a very special case in field theory.(no angular momentum,
no concentration of radial wave function at spatial origin,..),so it is 
of interest to know the results for the $2+1$ dimensional theory.
The tree diagram considerations which suggest large $1\rightarrow n$
amplitudes in
$\phi^4$ are mainly combinatoric, and hold as well in three dimensions,
as in two (or four). Furthermore, the $\phi^4$ theory is superrenormalizable
in three dimensions as well as in two. This is an essential element of
the analysis presented below. (In four dimensions one has to  deal
with the presumed triviality of the theory and a cutoff. In less than four
dimensions, the existence of the theory as the analytic continuation of the 
continuum limit of the Euclidean lattice theory is established.
\cite{GJ:qphy,FF:rwqft}. )

   To deal with large three dimensional lattices
we built a Beowulf parallel supercomputer with sixteen CPUs. We also 
used a Message Passing Interface(MPI) function call to parallelize our           simulation
program. We are able to achieve overall 7.4 Gflops at High Performance 
Computing Linpack Benchmark(HPL). The theoretical peak performance for sixteen
CPUs is 12 Gflops. Compared with a 16 Gflops CRAY C90 at 16 CPUs, our parallel
cluster has much better price/performance ratio.
(For details of the construction and performance of this Beowulf computer,
\cite{yyc:th}).

\section{\label{sec:level2}\protect\\Spectral function sum rules and bound
on high multiplicity production processes}


We start with some standard definitions to set the notation and display
the ingredients used in the analysis below. In d spacetime dimensions,
the canonical (unrenormalized) scalar field satisfies canonical 
commutation rules
\bdm
  [\dot{\phi}(x),\phi(0)]_{x_0 =0}=-i\delta^{d-1}(\vec{x}) \label{CCR}
\edm

The Wightman two-point function and its spectral decomposition are
\bdm
 \langle 0|\phi(x)\phi(0)|0\rangle =\int\;d\kappa^2\rho(\kappa^2)
 \Delta^{(+)}(x;\kappa^2)
\edm

\bdm
  \Delta^{(+)}(x;m^2)= \int(dq)e^{-iqx}\Theta(q_0)2\pi\delta(q^2-m^2)
\edm
\bdm
    \left( (dq)=\frac{d^d q}{(2\pi)^d} \right)  
\edm
\bea
  2\pi \rho(p^2) & = & \int(d^d x) e^{ipx} \langle 0|\phi(x)\phi(0)|0\rangle \\
   & = &(2\pi)^d {\mathbf \sum}_n\delta^{(d)}(p-p_n) 
 \langle 0|\phi(0)|n\rangle\langle n|\phi(0)|0\rangle \label{pos}
\eea
The states $|n \rangle $ are members of a complete set of in-,or out- states,
and are labeled as $|q_1,\ldots,q_n\rangle$ with
\bdm
 p_n^{\mu}=\sum_{a=1}^n q_a^{\mu}, \hspace{1in} \omega_{(a)}=\sqrt{\vec q_{(a)}^2
+m^2} 
\edm
The boldface sum over n implicitly includes some relativistic kinematical
factors relating
it to the relativistic invariant phase space integration.

\bdm
 {\mathbf \sum}_n|n\rangle \langle n| = |0\rangle\langle 0|+
   \sum_{n=1}^{\infty}\frac{1}{n!}\prod_{a=1}^n \left(\int\frac{d^{d-1}q_{(a)}}
  {(2\pi)^{d-1} 2\omega_{(a)}}\right)|q_1\ldots q_N \rangle \langle q_1\ldots    q_n|
\edm
Relativistic invariant phase space is simply related to this
\bdm
   (2 \pi)^d\delta^d (p-p_n)
    \frac{1}{n!}\prod_{a=2}^n \left(\int\frac{d^{d-1}q_{(a)}}
  {(2\pi)^{d-1} 2\omega_{(a)}}\right) =\int\;d\Phi_n(E)
\edm
where $E$ is the total cm energy $E=\sum_{a=1}^n q_{(a)}^0$

   Separate out the single particle contribution to the spectral function.
\bdm
  \rho(p^2)=Z\delta(p^2-m^2)+\hat{\rho}(p^2) \label{norm}
\edm
and
\bdm
  \langle q|\phi(0)|0\rangle = \sqrt{Z}
\edm
The CCR (\ref{CCR}) imply the sum rule
\bdm
 1 = \int d \kappa^2\rho(\kappa^2)
\edm
The integral is convergent in less than four space-time dimensions.
Together with the positivity of the spectral function (\ref{pos}),these
equations imply 
\bdm
   0\leq Z \leq 1
\edm
For free field, $Z=1$, and the field connects vacuum to single particle
state only. 

The renormalized field $\phi\hspace{-1.2mm}/(x)$ is defined by the normalization
condition
\bdm 
    \langle q|\phir(0)|0\rangle = 1
\edm
Thus $\phi(x)=\sqrt{Z}\phir(x)$.

For the multiparticle states (\ref{pos}),it is an exercise with the LSZ
reduction formulas to obtain
\bdm
  \langle q_1\ldots q_n|\phi(0)|0\rangle = \frac{1}{Z^{n/2}}G_F(p_n^2)Z^n
  \underline{\tau}_{n+1}(q_1\ldots q_n)
\edm
Here $\tau_{n+1}(q_1\ldots q_n)$ is the Fourier transform of
the vacuum expectation value of the T-product of n+1 $\phi$ fields,with the
energy-momentum conserving delta function left off. The underline indicates
that the external lines are all amputated ($q_1 \ldots q_n$ are all on-shell.)
$G_F$ is the complete unrenormalized time-ordered two-point function. 

The renormalization of all these quantities is
\bdm
  \rho = Z \rhor, \hspace{1in} G_F = Z \GFr, \hspace{1in}
    \underline{\tau}_{(n+1)}= \left(\frac{1}{Z^{\frac{n+1}{2}}}\right)            \underline{\taur}_{(n+1)}
\edm
 Then (\ref{pos}) becomes
\bdm
  2\pi\rhor_{(n)} = \int d\Phi_n(E)|\GFr(p_n^2)|^2 |\underline{\taur}_{(n+1)}
(q_1\ldots q_n)|^2
\edm
The $\GFr$ depends only on the total four-momentum $p_n$ (E in the c.m.)which
is fixed in the n particle phase space integral, so it can be divided out.
Divide also by particle density factor $2 E$
\bdm
 \frac{\pi}{E}\frac{\rhor_{(n)}(p_n^2)}{|\GFr(p_n^2)|^2} = \frac{1}{2 E}     \int d\Phi_n(E) |\underline{\taur}_{(n+1)}(q_1\ldots q_n)|^2  \\
 = \Gamma_{(n)}(E)
\edm
which is the decay rate for one variable mass $(E)$ $\phi$ particle to decay     into
n on-shell $\phi$ particles. The totally inclusive rate is
\bdm
  \Gamma(E)= \sum_n \Gamma_{(n)}(E)
\edm

One can think of the single timelike $\phi$ particle as formed by the
annihilation of an $f\bar{f}$ pair, coupled weakly to the $\phi$ field.
Then up to trivial kinematic factors, this is the analog of the QCD ratio
R of hadron production to $\mu^+\mu^-$ production in $ e^+ e^-$ annihlation.
We do not write a slash for $\Gamma$.There is no reason to introduce a
bare rate, so we simply write $\Gamma$ with no slash for the physical quantity.

  We now relate this rate to the inverse $\phi$ two-point function. (This
has a simple diagramatic representation. The probability for one timelike $\phi$
particle to go to many on-shell $\phi$ particles is given by the Cutkosky rules
applied to the cuts of the $\phi$ self-energy diagrams). 
The spectral representation of the Feynman function follows from (2)
\bdm
  G_F(p^2) = \frac{Z}{p^2-m^2}+\int\;d\kappa^2\frac{\hat{\rho}(\kappa^2)}        {p^2-\kappa^2+
i\epsilon}
\edm
The imaginary part is 
\bdm
  \Im G_F(p^2) = -\pi \hat{\rho}(p^2)
\edm
The imaginary part of $G_F^{-1}$ is
\bdm
  \Im G_F^{-1}(p^2)= -\frac{-\Im G_F(p^2)}{|G_F(p^2)|^2} 
  = \frac{\pi \hat{\rho}(p^2)}{|G_F(p^2)|^2} =\frac{E}{Z}\Gamma(E)
\edm
  From all the information about $G_F$, we can write a dispersion integral
for $G_F^{-1}$  \cite{Schw:B59}
\bdm
 G_F^{-1}(p^2) = (p^2-m^2)(1+\int\;d\kappa^2\frac{\gamma(\kappa^2)}{(\kappa^2-m^2)(\kappa^2-p^2-i\epsilon)} )
\edm
with
\bdm
 \Im G_F^{-1}(p^2) = \pi\gamma(p^2) = \frac{E}{Z}\Gamma(E),\linebreak
   \hspace*{2in} (p^2=E^2)
\edm
We note that all of the properties of $G_F$ enumerated so far are
consistent with the possible existence of zeros of $G_F$ on the real
axis. If $G_F$ has no additional poles, only one zero is possible, on
the real axis in the gap between the pole and the start of the continuum.
In any case, one can show that if such zero(s) of $G_F$ exist, positivity
of the spectral function is sufficient to show that the bound
derived below is only strengthened \cite{yyc:th}

The analytic continuation of the two-point function is immediate
\bdm
  p^0 \rightarrow ip_4,\hspace{.5in} p_M^2= -p_E^2, \hspace{.5in} G_F(p_M^2)
   = -G_E(p_E^2)
\edm
\bdm
   G_E^{-1}(p^2) = (p^2+m^2)(1+\int\;d\kappa^2\frac{\gamma(\kappa^2)}{(\kappa^2 - m^2)(\kappa^2+p^2)})
\edm
 In this and subsequent equations in this section $p^2$ is Euclidean $p_E^2$
The desired sum rules are now obtained by expanding both sides of this 
equation in powers of $p^2$ and equating the coeficients. The expansion
of the left hand side defines the quantities $Z^{\prime},m^{\prime^2}$
\bdm
   G_E^{-1}(p^2) =\frac{1}{Z^{\prime}}(m^{\prime^2} + p^2 +\ldots)  \label{m'Z'}
\edm
 while the right hand side is
\bdm
   G_E^{-1}(p^2) = m^2(1+\int\;d \kappa^2\frac{\gamma(\kappa^2)}{(\kappa^2 -m^2)
\kappa^2}) +p^2(1+\int\;d \kappa^2\frac{\gamma(\kappa^2)}{\kappa^4}) +\ldots 
\edm
Matching coeficientss of $p^2$ gives
\bdm
 \frac{1}{Z^{\prime}}= 1+ \int\;d \kappa^2\frac{\gamma(\kappa^2)}{\kappa^4}
 = 1+\frac{1}{\pi Z}\int\;d \kappa^2\frac{\kappa \Gamma(\kappa^2)}
{\kappa^4})
 = 1+\frac{2}{\pi Z}\int\;dE\frac{\Gamma(E)}{E^2}
\edm
Finally
\bdm
 \frac{2}{\pi}\int\;dE\frac{\Gamma(E)}{E^2} =Z(\frac{1}{Z^{\prime}}-1)            \leq \frac{1}{Z^{\prime}}-1   \label{inclus}
\edm
If the probability for production of high multiplicity states of
c.m. energy E turns on exponentially at some critical energy $E_{\star}$,
the integral of $\Gamma(E)/E^2$ will turn on exponentially as the range
of integration over E extends beyond $E_{\star}$. But by above, this integral is
bounded by $\frac{1}{Z^{\prime}}-1$, which can be extrapolated from lattice 
MC calculation of the slope of the inverse of the  Fourier Transform of
the two-point correlation.
 
\section{\label{sec:level3}\protect\\Critical energy and lattice momenta}

  According to the papers cited in the introduction, there may exist a critical
energy $E_{\star}$ ($\sim n\;m/g$) at which the probability for production 
of high multiplicity final states suddenly begins exponential growth. We
require an estimate of that critical energy and its relation to the 
momentum of the lattice propagator. Repeat the spectral representation
for $G_E(p^2)$ ($p^2$ is the Euclidean momentum squared).
\bdm
  G_E(p^2)= \frac{Z}{p^2+m^2} +\int\;d\kappa^2\frac{\hat{\rho}(\kappa^2)}        {p^2 + \kappa^2} 
\edm
$\kappa^2$ is the square of the c.m. energy of the state $|n\rangle$ 
contributing to the spectral function $\hat{\rho}$ (section two). For large
$p^2$,the
(convergent) integral is dominated by $\kappa^2 \leq p^2$. Then for the
integral to be sensitive to dramatic behavior for $\kappa^2\approx E^2_{\star}$
requires $p^2 \geq E_{\star}^2$. But the relevent $p$ is the Euclidean
lattice momentum which lies in the Brillouin zone $-\pi/2\leq p_{\mu}\leq
\pi/2$, or (lattice) $p^2\leq d(\pi /a)^2$. Combining these observations
leads to the desideratum
\bdm
  d\left(\frac{\pi}{a}\right)^2 > \Es^2
\edm
Take the square root and divide by mass $m$
\bdm
  \Es /m < \sqrt{d}\pi\xi_L    \label{Escond}
\edm
where $1/ma=\xi_L$ is the correlation length in lattice units.

  The estimation of $E_{\star}$ for the tree diagrams of $\phi^4$ in three
plus one dimensions has been refined in the papers cited in the introduction.
We have adapted these calculations to two plus one dimensions and find
\cite{yyc:th} 
\bdm
  \frac{\Es}{m}= 74\pi \frac{m}{\lambda}   \label{Es}
\edm 
Substituting this result into (\ref{Escond}) we obtain
\bdm
  74 \pi \frac{m}{\lambda} < \sqrt{d}\pi\xi_L
\edm

There is also a pair of general conditions for extrapolating results obtained
from a calculation on a lattice to results in continuum limit: The correlation
length should be much greater than the lattice spacing and much less than
the linear lattice size. Thus
\bdm
    1 << \frac{74}{\sqrt{3}}\frac{m}{\lambda} < \xi_L << N  \label{condition1}
\edm
Note that for fixed $\frac{\lambda}{m}$ of order one, these conditions
can all be satisfied by a large enough lattice (N of order one or 
two hundred).

\section{\label{sec:level4}\protect\\Lattice simulation and continuum limit}

  The Monte Carlo simulation for the scalar field on the lattice is standard.
The physical field, masses, and coupling constant are dimensionful quantities.When put on a lattice the lattice spacing a can be scaled out and the 
dimensionless action written in terms of dimensionless lattice fields and
mass and coupling constant. The distinction becomes important in taking 
the continuum limit. In three (Euclidean) dimensions the scalings are
\bdm
  \phi = \frac{1}{\sqrt{a}}\phi_L, \hspace{1in} \mu_0^2=\frac{1}{a^2}
    \mu_{0_L}^2, \hspace{1in}, \lambda= \frac{1}{a}\lambda_L  \label{latp}
\edm 
The lattice action is
\bdm
 S=\sum_{\vec{n}} \left\{ \frac12\sum_{\nu}
\left[ \phi_L(\vec{n}+\vec{e}_{(\nu)}) - \phi_L(\vec{n}) \right]^2
+\frac12\mu^2_{0_L}\phi_L(\vec{n})^2+\frac14\lambda_L\phi_L(\vec{n})^4 \right\}
  \label{Lag} 
\edm

We start with a standard Metropolis update algorithm, and with each 
sweep of the lattice measure the expectation of $\phi$ and the two-point
correlation. (For this discussion of the lattice calculation all quantities   are in lattice units. We supress the subscript L until we come back to the 
continuum limit). We take the lattice Fourier transform of $\phi$ and          compute
the output lattice propagator
\bdm
        G(k)=\langle \phi(k)\phi(-k)\rangle
\edm
and plot the inverse of this against the inverse of the free lattice           propagator
\bdm
   G_0^{-1}(k)=\hat{k}^2 + m^2
\edm
with
\bdm
  \hat{k}^2 = 4\sum_{\nu}\sin^2(k_{\nu}/2)   \label{mflp}
\edm
This is the inverse of the massless free lattice propagator.

Z' and m' are obtained from the plot of $G^{-!}(\hat{k}^2)$ against the 
inverse of the free massless lattice propagator.(\ref{m'Z'})

\begin{figure}[htbp]
\begin{center}
\leavevmode
\epsfxsize 6in
\epsfbox{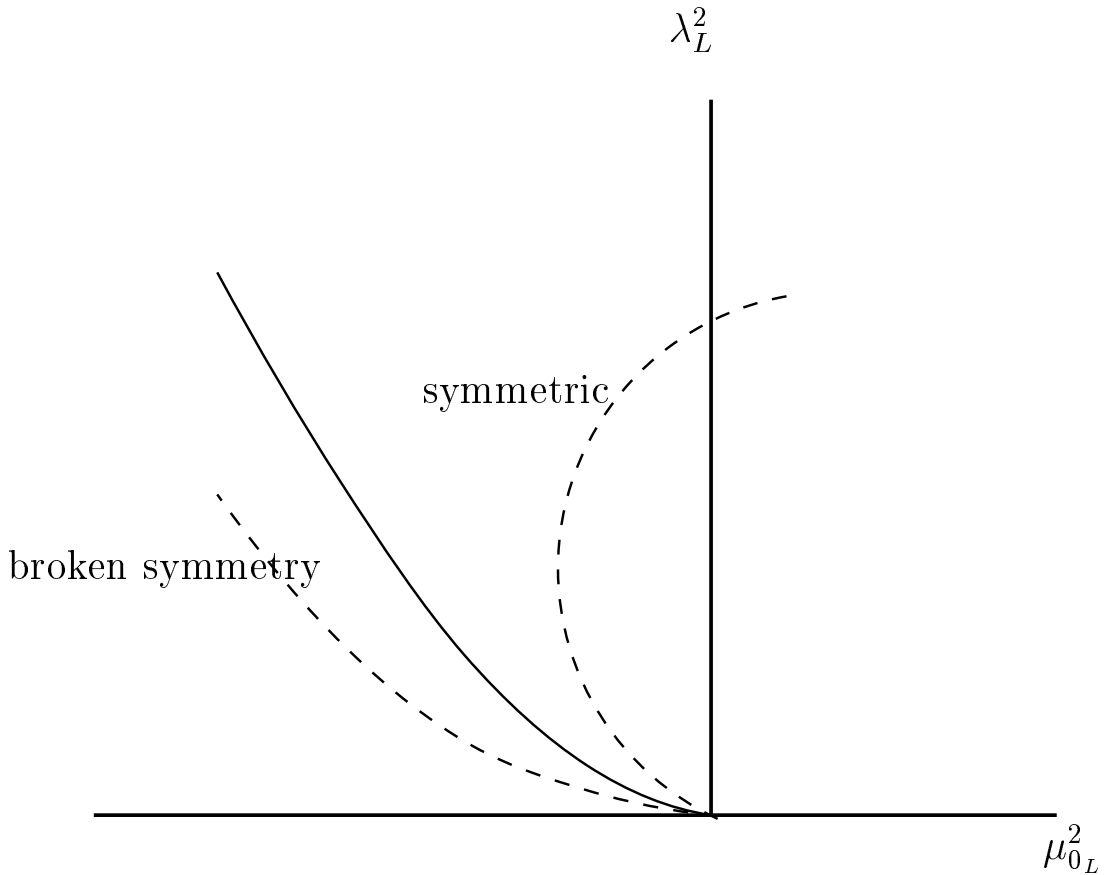}
\caption{The phase diagram for the three-dimensional, single component scalar
$\phi^4$ theory. The solid line separates the symmetric phase from the broken
symmetry phase. 
The dashed lines are lines of constant physics.}
\label{fig:fasediag}
\end{center}
\end{figure}

 As we approach the contiuum limit we are also approaching the critical
line fig (\ref{fig:fasediag}), and the simple Metropolis update begins to suffer  
 critical slowing down.
 As one approaches the critical line, both the
correlation length and autocorrelation time diverge. The relation of the
correlation length $\xi$ and autocorrelation time $\tau$ is\cite{HH:dynz,Lt:QFT}
\begin{equation}
\tau \sim \xi^z \sim L^d
\end{equation}
where $z$ is the dynamic exponent and $\tau$ is measured in
number of sweeps.
In a system                                                              
of finite size, the correlation length never really diverges. Cutting it
off at the size of the lattice
, the autocorrelation time increases as the lattice size increases as
\begin{equation}
\tau \sim L^z
\end{equation}
when close to critical. Depending on the value of z, one can require 
a large amount of CPU time to generate a fixed number of independent 
data as one approaches critical on large lattices.For straight Metropolis
update, $z$ is close to two. See Table I.

To overcome the critical slowing down, one must define appropriate nonlocal
(or collective) variables and a new dynamics for driving them. Progress has
been made with some nonlocal algorithms for both discrete spin models and
continuum fields. Swendsen and Wang\cite{SW:cls} have used the Fortuin-Kastelyn
\cite{FK:pha} percolation map for the Potts model to define collective
coordinates that allow domains to be inverted with zero free energy cost.
Some modifications were proposed by Wolff\cite{Wf:prl23} 

The algorithm for the $\phi^4$ field theory is a little more
complicated. Brower and Tamayo\cite{BT:phi4cls}  proposed an algorithm for
$\phi^4$ field theory based on the Swendsen and Wang algorithm. Here we will
follow Brower and Tamayo's method but switch to the Wolff algorithm. Table
\ref{dynexp}
compares the dynamic exponent values for the Ising model in different dimensions with different algorithms.
 One can see that the Wolff algorithm is more effective
than the Swendsen and Wang algorithm in three dimension.

\begin{table}[tbp]
\caption{Comparison of the values of the dynamic exponent $z$ for
different algorithms for Ising model in various dimensions\cite{dy:z}.}
\label{dynexp}
\begin{center}
\begin{tabular}{|c|c|c|c|}
\hline
dimension d & Metropolis & Wolff & Swendsen-Wang \\
\hline
2 & 2.167 $\pm$ 0.001 & 0.25 $\pm$ 0.01 &  0.25 $\pm$ 0.01 \\
3 & 2.02 $\pm$ 0.02 & 0.33 $\pm$ 0.01 & 0.54 $\pm$ 0.02 \\
4 & - & 0.25 $\pm$ 0.01 & 0.86 $\pm$ 0.02 \\
\hline
\end{tabular}
\end{center}
\end{table}

In our simulation, the
update algorithm consists  of two parts:{\bf (1)} a conventional
Metropolis Monte Carlo update for the $\phi(x)$ field and {\bf(2)} Wolff cluster identification and flipping.  Identification of a cluster of $\phi$ field,
requires introduction of a discrete variable $s_{\vec{n}}$ 

\begin{equation}
\phi(\vec{n}) = s_{\vec{n}} |\phi(\vec{n})|
\end{equation}

where  $s_{\vec{n}}=\pm 1$.

\begin{figure}[htbp]
\begin{center}
\leavevmode
\epsfxsize 4in
\epsfbox{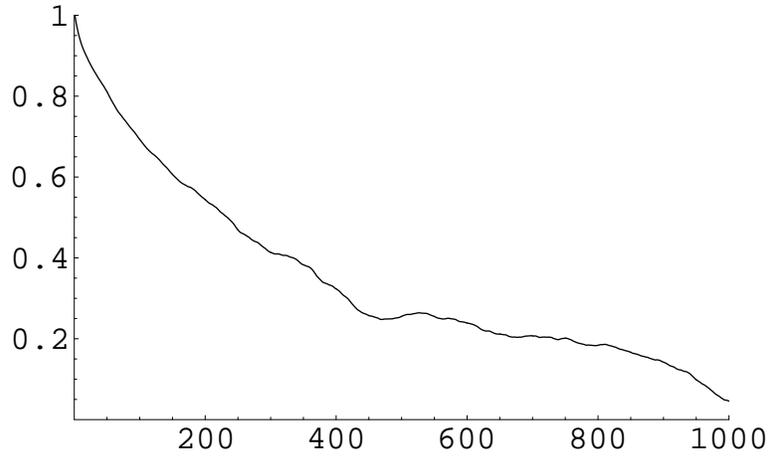}
\vskip 2cm
\epsfxsize 4in
\epsfbox{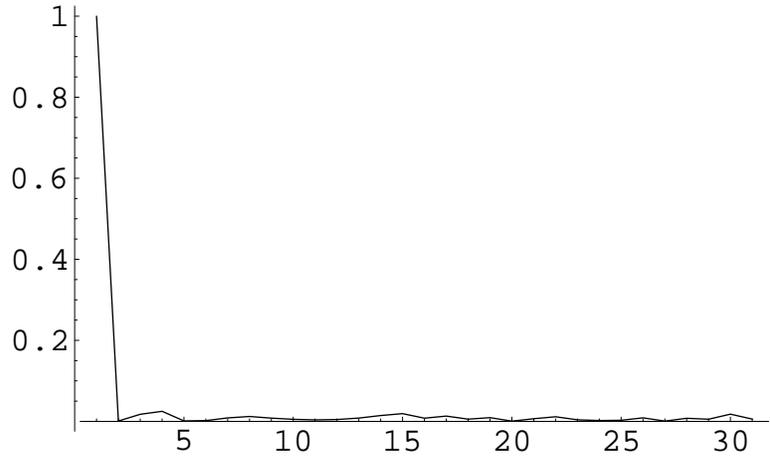}
\caption{The autocorrelation for the $\langle \phi \rangle$ with and
without the cluster update algorithm. The $e^{-1}$ point for the
one without the cluster update algorithm is at about 400(top).
The autocorrelation length for the one with cluster update algorithm is about
2 only(bottom).}
\label{fig:corr}
\end{center}
\end{figure}

For details of the cluster algorithm see \cite{yyc:th} Here we only
present some illustration of its effectivness and correctness.In fig(\ref{fig:corr} )
we show plots of autocorrelation in measurements of $\langle\phi\rangle$
for one set of input parameters, with and without the cluster algorithm.
We see an improvement by a factor of roughly 200 in the number of sweeps 
required to reach $e^-1$ fall off.

 There is a price to be paid for this 
improvement in generation of uncorrelated equilibrium configurations. That
is in the actual measurement of $\langle\phi\rangle$ itself. With only
Metropolis update, on large lattices, there are very many sweeps between
tunneling, and measurement of $\langle\phi\rangle$ is trivial. The cluster
update changes sign of $\langle\phi\rangle$ frequently and it is harder to
dig out its infinite volume (no tunneling) limit.
\clearpage

\begin{figure}[tbp]
\begin{center}
\leavevmode
\epsfxsize 5in
\epsfbox{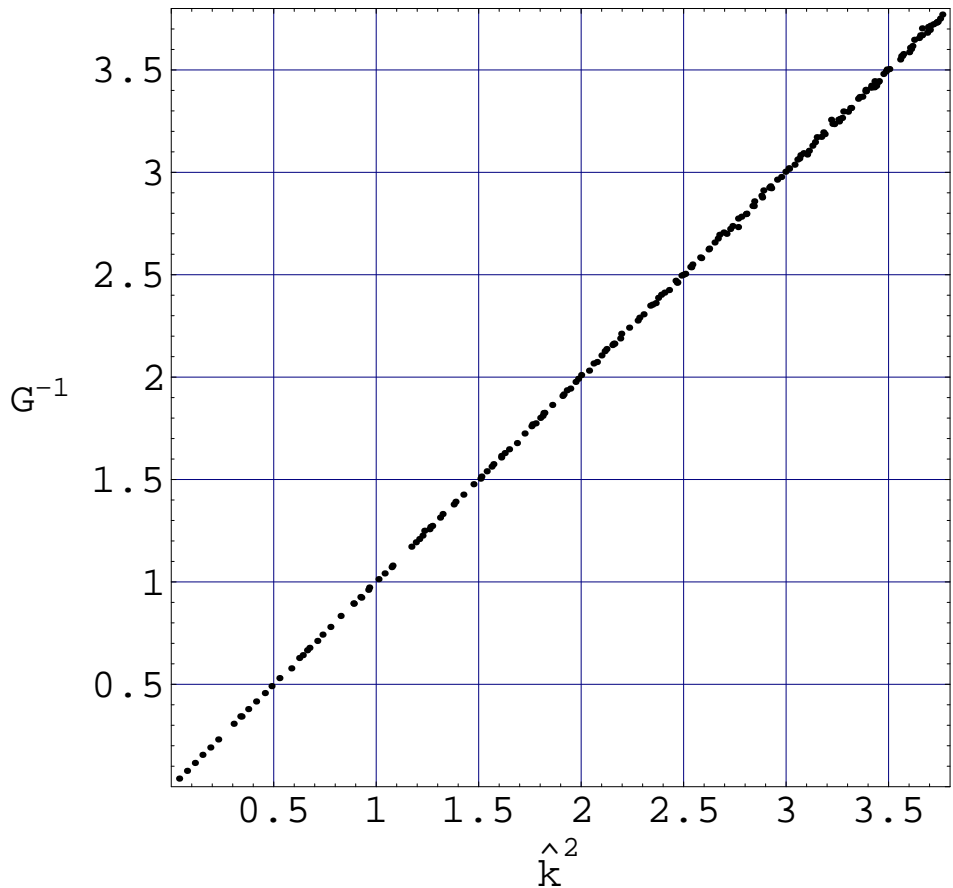}
\caption{Result of Monte Carlo evaluation of $G^{-1}(k)$ for $\lambda=0, 
\mu_0^2=0.001$ plotted against exact analytic massless propagator
$\hat{k}^2 = 4\sum_{\nu} \sin^2(k_{\nu}/2)$ 
on a $32^3$ lattice. Error bars on Monte Carlo results are too
small to show up on this plot.}
\label{figure:freeprop}
\end{center}
\end{figure}

 An important check is
to do the MC simulation for the coupling constant $\lambda$ equal to zero.
For this case we know the exact analytic (free field) solution. In particular,
$m^{\prime^2}$ is equal to input bare mass squared (necessarily positive) and
$Z^{\prime}=Z =1$). This is shown in fig\ref{figure:freeprop} where $G^{-1}$     from 
the simulation (with $\lambda=0$)is ploted against the free massless inverse
propagator $(\hat{k}^2)$. The result is a straight line of unit slope, 
offset by the mass squared ($0.001$ and not visible in this example).   

Although the physics is entirely dependent on $\lambda \neq 0$, for the MC 
simulation program and the program for analysis of the MC data, zero is just
one possible value of $\lambda$. So the above result does check a substantial
part of the simulation and data analysis programs.

For$\lambda \neq 0$, we have also, for a few sets of input parameters,paid
the price in longer CPU time required when the cluster algorithm is turned
off, and checked that the results $(m^{\prime},1/Z^{\prime})$ are the same,
within small statistical errors, with or without the cluster algorithm.
    
We will return to another test, the Ising limit, after our discussion of 
the phase structure and the continuum limit. 

There are both general theoretical and heueristic and practical
considerations involved in extracting real world results from the
Euclidean lattice results. The general theoretical underpinning comes from
the program of constructive field theory which has provided the results that
for $d<4$, the $\phi^4_d \;$ Quantum Field Theory can be obtained as the 
analytic continuation of the Euclidean $\phi^4_d$ field theory, which
in turn may be obtained as the continuum limit of a lattice $\phi^4_d$
theory. This program is described in detail in two long and mathematical
monographs \cite{GJ:qphy},\cite{FF:rwqft}. A simple example is again 
provided by the free theory $(\lambda=0)$ which is completely determined
by the two-point function . When the lattice propagator (\ref{mflp}) is            rewritten in terms of physical dimensionful quantities with the 
lattice spacing $a$ inserted where required to give the propagator its            correct physical dimension, the limit $a\rightarrow 0$ is simply seen 
to be the free Euclidean propagator. 
However, in the general case, the output from the MC simulation is 
numerical, and the limit has to be done by extrapolation. To see what
is involved, we consider the phase diagram fig (\ref{fig:fasediag}).

 The $\phi^4$ theory on a lattice of lattice spacing $a$ has two dimensionless
parameters, $\mu_{0_L}^2,\lambda_L$ (\ref{latp}). Since $\lambda$ is finite
and $\mu_0^2$ is only linearly divergent ($\sim 1/a$), both of these 
dimensionless parameters go to zero in the limit $a\rightarrow 0$.Thus the 
point at the origin in fig(\ref{fig:fasediag}) corresponds to the Euclidean field     theory. So start
on dimensionless lattice ($N^3$)
at some point in $\mu_{0_L}^2,\lambda_L$ plane. Choice of parameters
is subject to requirement that output $\xi_L^{\prime}=1/m_L^{\prime}$             satisfy 
\bdm
 1 \ll\xi_L^{\prime}\ll N   \label{condition2}
\edm
 This is an obvious condition for lattice spacing
and volume effects to be small. To proceed toward the origin, go to lattice
$N_2 = 2N$, and adjust input $\mu_{0_L}^2,\lambda_L$ to get output
$\xi_{L_2}^{\prime}=2\xi_L^{\prime}$.This corresponds to keeping same 
physical volume and correlation length while reducing lattice spacing
to half of starting value. A sequence of these steps corresponds to
one of the dashed lines in fig(\ref{fig:fasediag}). Note that at each 
step $\xi_L^{\prime}$ is increasing (doubling) so the first inequality in
(\ref{condition2}) is increasingly better satisfied (The second remains the same).

The $\phi^4$ field theory ($a \rightarrow 0$) is specified by a single finite 
dimensionless constant, which we may take to be $\lambda/m'$; so ``constant
physics'' is implemented by fixed value of $\lambda/m'$.Also,
 the two other finite dimensionless quantities we can measure,{$1/Z'$},
and $\langle\phi\rangle /\sqrt{m'}$,are functions of $\lambda/m'$ in the 
limit of $a \rightarrow 0$. As the origin
is approached along one of the dashed lines, the values of these quantities
should approach their continuum values. In practice, When the change in values
in succesive steps is less than the estimated error, we accept those values.

 We briefly discuss the relation of this work to the Ising model of          Statistical
Mechanics. This is in itself of some interest, and it also provides another
check of our MC simulations. The EFT limit is reached by following one of the
dashed lines into the origin ($\lambda_L\rightarrow 0$). The Ising model limit
is approached by following a dashed line (in broken symmetry phase) out to 
infinity ($\lambda_L\rightarrow \infty$).The Euclidean lattice Lagrangian
(\ref{Lag}) can be rewritten as

\begin{equation}
{\cal L} = -\sum_{\langle\vec{i},\vec{j}\rangle}\phi(\vec{i}) \phi(\vec{j}) +
\sum_{\vec{i}}\frac{\lambda}4 \left( \phi^2(\vec{i}) -
\frac{-\mu_0^2+2d}{\lambda} \right)^2  \label{eq:ftising}
\end{equation}
All quantities here are in lattice units (subscript L suppresed).The sum over
angle bracket is standard notation for sum over nearest neighbors, each pair
counted once. An irelevent constant has been added.
 Compare with the Hamiltonian of Ising model
\begin{equation}
{\cal H} = -\beta \sum_{\langle\vec{i},\vec{j}\rangle} \sigma_{\vec{i}}\sigma_{\vec{j}}  \label{IsingH},
\end{equation}
The limit $\lambda \rightarrow \infty$ will exponentially suppress the    second term in
equation (\ref{eq:ftising}), and convert the continuous field variable to a 
spin variable.
\begin{equation}
\lim_{\lambda\rightarrow \infty} \phi(\vec{i}) =\pm \sqrt{\frac{-\mu_0^2+2d}{\lambda}}\equiv \sqrt{\beta}\sigma_{\vec{i}}
\end{equation}
with $\sigma_{\vec{i}}=\pm 1$ and
\bdm
 \beta=\frac{-\mu_0^2+2 d}{\lambda}
\edm

To compare with the Ising limit,
we do the simulation with the $\lambda \gg 1$  and
calculate the VEV $\langle \phi \rangle$ for different $\mu^2_0$. We start
in the symmetric phase with one $\lambda$ value and vary the value of $\mu^2_0$
until the VEV is no longer zero. In our simulation, when $\lambda \gg 1$, we
get a very sharp transition from the symmetric phase to broken symmetry phase.
So, it is very easy to identify the critical $\mu^2_{0_c}$.
Table \ref{isinglim} shows the $\beta_c$ from the simulation at $32^3$
with different
$\lambda$. The third row shows that when $\lambda=1000$, the $\beta_c$ from
our simulation with a continuous field variable is very close to the (numerically) known value 0.2217... for the 3d Ising Model  This give us additional           confidence in our MC simulation.

\begin{table}[tbp]
\caption{The Monte Carlo simulation of $32^3$ $\phi^4_3$ theory at Ising
limit($\lambda \rightarrow \infty$). The table shows the critical temperature
$\beta_c$ at different $\lambda$ values.}
\label{isinglim}
\begin{center}
\begin{tabular}{|c|c|c|c|}
\hline
$\lambda$ & $\mu^2_{0_c}$ & $-\frac{\mu^2_{0_c}}{\lambda}$ & $\beta_c$ \\
\hline
100 & -31 & .31 & .25 \\
300 & -75 & .25 & .23 \\
1000 & -228 & .228 & .222 \\
\hline
\end{tabular}
\end{center}
\end{table}.

\section{\label{sec:level5}\protect\\Analysis and results}

   A fortran program mplementing the Metropolis and cluster update algorithms  described above provides the equilibrium lattice configurations i.e. the sets
 of values
 $\phi(\vec{n})_t$ for t from 1 to $N_t$ where $t$ denotes the t'th sweep  
through the lattice and$N_t$ is the 
number of sweeps through the lattice after thermalization. For large lattices
with $N_t$ less than the time before tunneling, and the cluster update 
turned off, the vev is

\bdm
  <\phi> = 1/N_t 1/N^3 \sum_1^{N_t}\sum_{\vec{n}}\phi(\vec{n})_t 
\edm

When the cluster update is turned on, the sign of the vev as defined above
changes frquently and $<\phi>$ is very small, even deep in the broken 
symmetry phase. So the vev is then defined as

\bdm
  <|\phi|> = 1/N_t 1/N^3 \sum_1^{N_t}|\sum_{\vec{n}} \phi(\vec{n})_t| 
\edm

  The program also does the lattice Fourier transform of $\phi$ and
constructs the k-space lattice propagator. This output is then input for
a Mathematica program which determines $m^{\prime^2}/Z^{\prime}$ and
$1/Z^{\prime}$ from fit of output lattice propagator 
\bdm
  G(\hat{k}^2)^{-1}=(1/Z^{\prime})(m^{\prime^2}+\hat{k}^2+\ldots)
\edm 
The estimated errors asociated with $1/Z^{\prime}$ and $m^{\prime}$
for fixed lattice (N) are statistical, originating from the Monte Carlo,
and systematic from extrapolation to k equal to zero in the fit.                (We can not use
the data point $k=0$ because it contains a volume singularity) 
The results are collected in Table \ref{tb:73}

\begin{table}[hbt]
\caption{The results of fit to data from the simulations on different size
lattices($N^3$),the input parameters $\mu^2_{0_L},\lambda_L$,and the output
 dimensionless coupling parameter $\lambda/m'$,and 
the output correlation length $\xi'_L=1/m'_L$,the slope $1/Z'$, and
the dimensionless ratio  $<|\phi|>/\sqrt{m'}$ .}   \label{tb:73}
\vskip 1cm
\begin{center}
\begin{tabular}{|c|c|c|c|c|c|c|}
\hline
$N^3$ & $\mu^2_{0_L}$ & $\lambda_L$ & $\frac{\lambda}{m'}$ & $ \xi'_L$   &      $\frac1{Z'}$ & $\frac{\langle \phi_0 \rangle}{\sqrt{m'}}$  \\
\hline
$32^3$ & -0.161 & 0.2  & 1.13(2) & 5.7(1) & 1.008(2) & 0.78(1)   \\
$64^3$ &  -0.078 & 0.1  & 1.13(4) & 11.3(4) & 1.012(3) & 0.80(2)   \\
$128^3$ &-0.0384 & 0.05  & 1.11(5) & 22.11(1.0) & 1.012(5) & 0.80(2)   \\
$256^3$ &-0.01905 & 0.025 & 1.13(8) & 45.4(3.0) & 1.010(8) & 0.79(3)   \\
\hline

\end{tabular}
\end{center}
\end{table}

\begin{figure}[htb]
\begin{center}
\vskip 1cm
\leavevmode
\epsfxsize 4.5in
\epsfbox{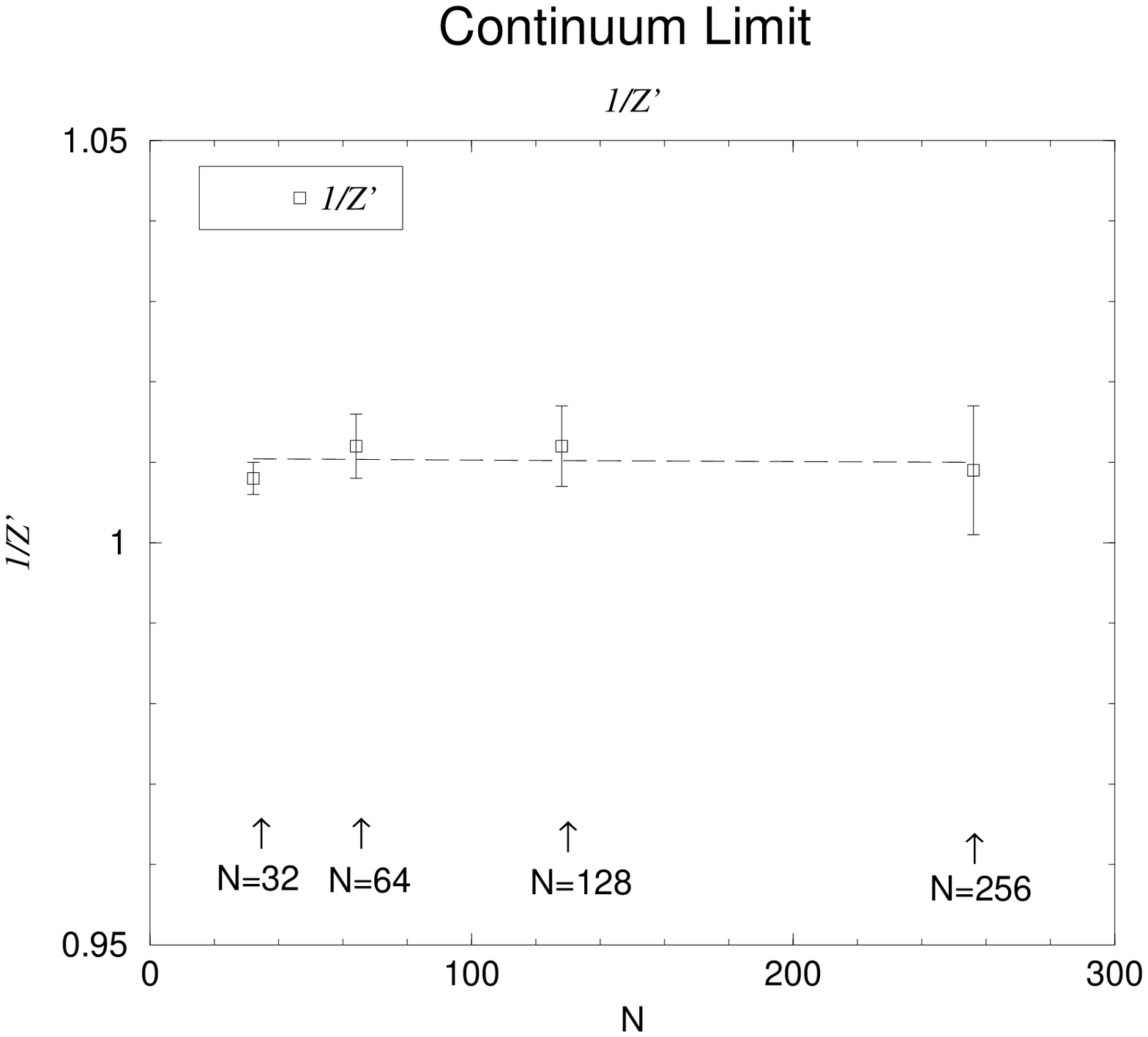}
\vskip .1cm
\caption{Continuum limit for  $\frac1{Z'}$. The
plot shows the insensitivity to lattice spacing for $\frac1{Z'}$.}
\label{fig:71}
\end{center}
\end{figure}

\begin{figure}[htb]
\begin{center}
\vskip 1cm
\leavevmode
\epsfxsize 4.5in
\epsfbox{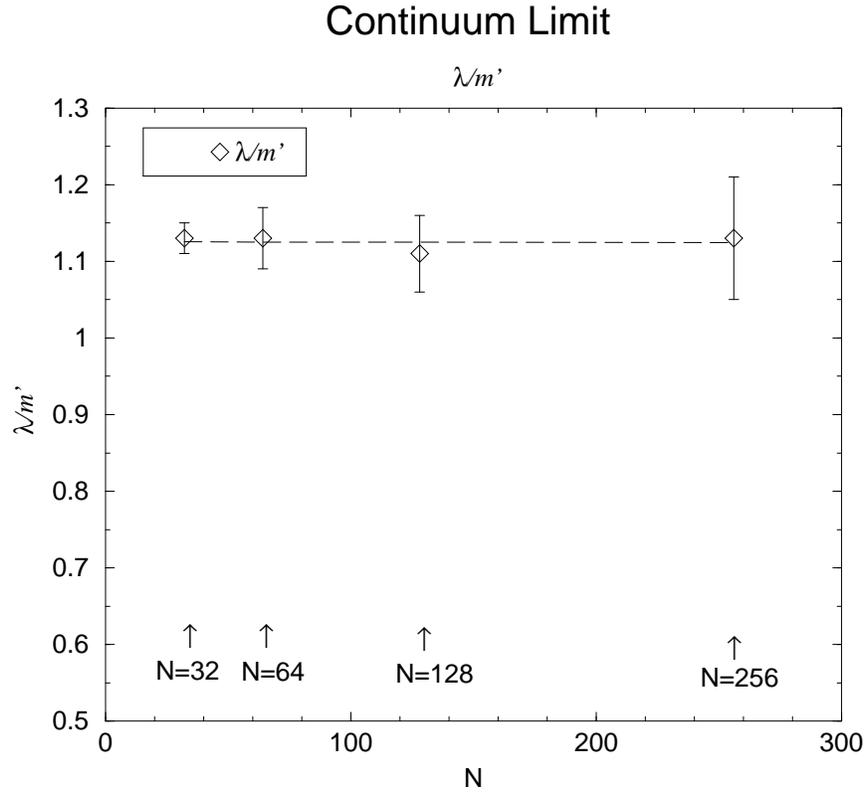}
\vskip .1cm
\caption{Continuum limit for the $\frac{\lambda}{m'}$. The
plot shows the insensitivity to lattice spacing for $\frac{\lambda}{m'}$.}
\label{fig:72}
\end{center}
\end{figure}

\begin{figure}[htb]
\begin{center}
\vskip 1cm
\leavevmode
\epsfxsize 4.5in
\epsfbox{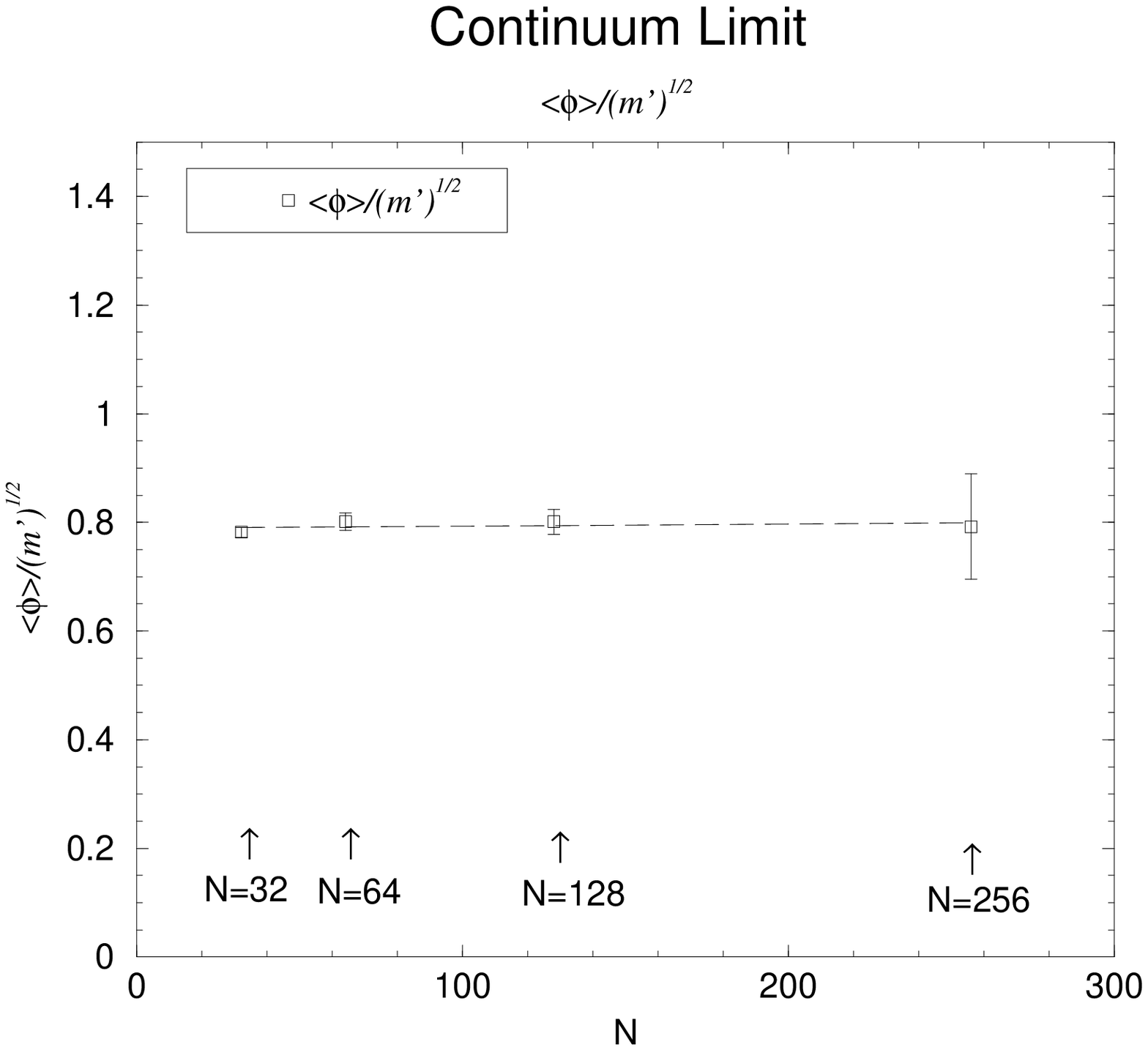}
\vskip .1cm
\caption{Continuum limit for the $\frac{\langle \phi_0 \rangle}{\sqrt{m'}}$. The
plot shows the insensitivity to lattice spacing for 
$\frac{\langle \phi_0 \rangle}{\sqrt{m'}}$.}
\label{fig:73}
\end{center}
\end{figure}

We can see in the Table the
implementation of the approach to the continuum limit by a sequence of
steps in which the physical correlation length (physical mass) is kept
fixed while the linear number of lattice points (N) and the correlation 
length in lattice units ($\xi'_L$) are both doubled. The ratio
$N/\xi'_L \;\;(L/\xi)\;$ is kept fixed ($\sim 5.6$).
 We also see that
$\lambda/m'$ keeps the same value within expected errors through these steps
as do the dimensionless quantities, $1/Z'$, $\langle\phi\rangle /\sqrt{m}$   
,consistent with their being functions of $\lambda/m'$ only. This suggests that 
these values are insensitive to lattice artifacts  and approaching the 
continuum limit. This is illustrated in figures (\ref{fig:71}),(\ref{fig:72}),  (\ref{fig:73}). 

\nopagebreak

\nopagebreak

\nopagebreak

  We also check the absence of finite volumn effects. We hold the input
parameters,$\mu_{0_L}^2,\lambda_L$,fixed and vary N. In Table \ref{tab:74} we    see
that when the linear size of the lattice is greater than three times the
output correlation length, the output quantities are unchanged under further
increase in N. Note that in Table \ref{tb:73} the ratio of N to  $\xi'_L$ is    kept
fixed at 5.6 as one moves along a dashed curve in fig(\ref{fig:fasediag})

\begin{table}[htbp]
\caption{The finite size effect with fixed input parameters $\mu^2_{0_L}=0.01905$,
$\lambda_L=0.025$.} \label{tab:74}
\vskip 1cm
\begin{center}
\begin{tabular}{|c|c|c|c|c|c|c|}
\hline
$\mu^2_{0_L}$ & $\lambda_L$ & $N^3$ & $\xi_L$ & $\frac{\lambda}{m'}$ & $\frac1{Z'}$ & $\frac{\langle \phi_0 \rangle}{\sqrt{m'}}$  \\
\hline
-0.01905 & 0.025 & $64^3$ & 41.5(6.5) & 1.04(16) & 1.002(3) & 0.78(14)  \\
-0.01905 & 0.025 & $128^3$ & 45.1(3.8) & 1.127(97) & 1.0011(35) & 0.77(10) \\
-0.01905 & 0.025 & $256^3$ & 45.4(3.0) & 1.13(8) & 1.010(8)  & 0.795(97)  \\
\hline
\end{tabular}
\end{center}
\end{table}

\clearpage

  The significance of these results is brought out in the qualitative behavior  of the
quantities in Table \ref{tb:75}. 

\begin{table}[htbp]
\caption{Size of lattice and condition for appearance of high multiplicity
 inelastic processes}  \label{tb:75}
\vskip 1cm
\begin{center}
\begin{tabular}{|c|c|c|c|c|}
\hline
 $1\;\;\;<<$ & $\frac{74}{\sqrt{3}}\frac{m'}{\lambda}\;\;\; <$ & $ \xi_L\;\;\;<$ & $ \;\;\;<\;N $ 
& $1/Z'-1$  \\
\hline
 1 & 37.8 & 5.7 & 32 & .008(2)  \\
 1 & 37.8 & 11.3 & 64 & .012(3) \\
 1 & 38.5 & 22.1 & 128 & .012(5) \\
 1 & 37.8 & 45.4 & 256 & .010(8) \\

\hline

\end{tabular}
\end{center}
\end{table}

The condition (\ref{condition1}),(\ref{condition2})

\bdm
    1 << \frac{74}{\sqrt{3}}\frac{m'}{\lambda} < \xi_L << N  \label{condition}
\edm

\nopagebreak{is not satisfied for the first three lattices, but
is clearly satisfied for the last (finest grained) lattice. And there is no
significant change in the value of $1/Z'-1$ in going from the first lattice  to
the last. We consider this to be strong nonperturbative evidence that there
is no exponential turn on of high multiplicity production processes for energies greater than some critical energy of order $n\frac{m}{\lambda}$}

Allthough the most significant feature of Table\ref{tb:75} is the absence of    large \textit {change}  
in $1/Z'-1$ as one passes into a region of parameter space in 
which the divergence of perturbation theory suggests the possibility of
exponential growth of high multiplicity processes; one can
   still ask: what would be a large value or a small value of $1/Z'-1$          in this context? What is
the significance of the single number $0.010$ ? In the
weak coupling regime, we can take low order perturbative effects to define
small. We have calculated $1/Z'-1$ through two-loop order, 
 in broken symmetry phase. \cite{yyc:th}.
We have chosen to specify the constructed $\phi^4_3$ theory by the 
output $\lambda/m'$. For the perturbation expansion parameter (g) we take
$\frac{\lambda}{8\pi m'}$   
 The result is
\bdm
 \frac1{Z'} = 1+\frac{\lambda}{8\pi m'}(3/4) +(\frac{\lambda}{8\pi m'})^2
(-1.83) + \ldots    \label{Z'}
\edm
For our particular lattice MC solution, $\lambda/m'=1.13$ ($g=.045$). This 
gives the perturbative value
\bdm
  1/Z'-1 = .0337 - .0037 +\ldots \simeq  .030
\edm
The lattice MC result from Table \ref{tab:74} is
$.010(8)$, which is close to consistent with the perturbative estimate.
 So any nonperturbative 
contribution is of the same order as or smaller than the perturbative estimate, which is small.

We can still ask: If there were important high multiplicity processes 
contributing to (\ref{inclus}), how would they show up in this analysis? 
From all the spectral equations in section II, we see that this would
show up as Z substantially less than one. In this case, the last step
in (\ref{inclus}) leads to an inefficient bound. So we go back to
\bdm
 \frac{2}{\pi}\int\;dE\frac{\Gamma(E)}{E^2} =Z(\frac{1}{Z^{\prime}}-1)            \leq \frac{1}{Z^{\prime}}-1
\edm
By manipulation of the equations of section II, one can show that 
$Z/Z'\leq 1$. Then on the righthand side of the equation
\bdm
 Z(\frac{1}{Z'}-1)=\frac{Z}{Z'}-Z \leq 1-Z \leq 1
\edm
So even if the decay were dominated by high multiplicity final states, the
integral of the inclusive rate is bounded by one.
\bdm
 \frac{2}{\pi}\int\;dE\frac{\Gamma(E)}{E^2} \leq 1
\edm
While in this limit we expect 1/Z' to be much greater than one i.e.
1/Z' -1 is a very inefficient bound in this case.

It is also interesting to compare the perturbative and nonperturbative 
calculations of $<|\phi|>/\sqrt{m'}$.The perturbative expression is
\bdm
 \sqrt{\frac{m'}{2\lambda}}(1 + \frac{\lambda}{8\pi m'}(39/8) + \ldots)
\edm
For $\lambda /m'= 1.13$ this is $0.81$. The lattice MC value from 
Table\ref{tab:74} is $0.795(95)\;$ ( or rounding $0.80(1)$ )

In conclusion: The absence of any significant change in the value of
$1/Z'-1$ from the lattice MC calculation as one moves from coarse grained
to fine grained lattices, and that the value is close to the two-loop 
perturbative result,  are strong evidence that there is no exponential          enhancement
of the rate for the high multiplicity processes $ 1 \rightarrow n$ for
$n \simeq 1/g$. We are not sensitive to the presence or absence of 
exponential suppression of rates for asymptotic values of E,n, which 
by definition are very small and only effect terms far out in the series
(\ref{Z'}).

\begin{acknowledgments}
 Y.Y.Charng is partially supported by the National Science Council
 of the Republic of China under the Grant No. of NSC-{\bf 90}-
 2811- {\bf 001-071} 
\end{acknowledgments}

\bibliography{basename of .bib file}

\end{document}